\documentclass[prb,aps,twocolumn,floatfix,superscriptaddress]{revtex4}

\usepackage{graphicx,graphics}

\begin{document}

\title{Structure and energetics of Si(111)-(5$\times$2)-Au}
\author{Steven C. Erwin}
\affiliation{Center for Computational Materials Science, Naval Research Laboratory, Washington, DC 20375, USA}
\author{Ingo Barke}
\affiliation{Institut f\"{u}r Physik, Universit\"{a}t Rostock, D-18051 Rostock, Germany}
\author{F. J. Himpsel}
\affiliation{Department of Physics, University of Wisconsin-Madison, Madison, WI 53706, USA}

\date{\today}

\begin{abstract}
  We propose a new structural model for the Si(111)-(5$\times$2)-Au
  reconstruction. The model incorporates a new experimental value of
  0.6 monolayer for the coverage of gold atoms, equivalent to six
  gold atoms per 5$\times$2 cell. Five main theoretical results, obtained
  from first-principles total-energy calculations, support the model.
  (1) In the presence of silicon adatoms the periodicity of the gold
  rows spontaneously doubles, in agreement with experiment. (2) The
  dependence of the surface energy on the adatom coverage indicates
  that a uniformly covered phase is unstable and will phase-separate
  into empty and covered regions, as observed experimentally. (3)
  Theoretical scanning tunneling microscopy images are in excellent
  agreement with experiment. (4) The calculated band structure is
  consistent with angle-resolved photoemission spectra; analysis
  of their correspondence allows the straightforward assignment
  of observed surface states to specific
  atoms. (5) The calculated activation barrier for diffusion of
  silicon adatoms along the row direction is in excellent agreement with
  the experimentally measured barrier.
  
\end{abstract}


\maketitle

\section{Introduction}

Forty years ago Bishop and Riviere first observed the (111) surface of
silicon to reconstruct, with fivefold periodicity, in the presence of
gold.  \cite{bishop_j_phys_d_appl_physics_1969a} Since that time the
Si(111)-(5$\times$2)-Au reconstruction has been widely studied, in
several hundred publications, as a prototype linear metallic chain
system in which the physics of one-dimensional metals is approximately
realized.\cite{hasegawa_j_phys_condens_matter_2000a,matsuda_j_phys_condens_matter_2007a}

These investigations have provided very substantial insights into
many aspects of Si(111)-(5$\times$2)-Au.\cite{barke_solid_state_comm_2007a,barke_appl_surf_sci_2007a}
Less successful have been the many attempts
to use the clues provided by experiment to construct a complete
structural model for this complicated reconstruction.  Beginning with
the early work of LeLay over a dozen different models have been
proposed.\cite{lelay_surf_sci_1977a,berman_phys_rev_b_1988a,
hasegawa_journal_of_vacuum_science__technology_a_1990a,
bauer_surf_sci_1991a,
schamper_phys_rev_b_1991a, seehofer_surf_sci_1995a,
marks_phys_rev_lett_1995a, plass_surf_sci_1997a,
omahony_phys_rev_b_1994a,omahony_surf_sci_1992a,
shibata_phys_rev_b_1998b,hasegawa_surf_sci_1996a,hasegawa_phys_rev_b_1996a,hasegawa_surf_sci_1996b,
erwin_phys_rev_lett_2003a,kang_surf_sci_2003a,
riikonen_phys_rev_b_2005a,ren_phys_rev_b_2007a,chuang_phys_rev_b_2008a}
All were eventually found to be inconsistent with the results of
scanning tunneling microscopy (STM), angle-resolved photoemission spectroscopy
(ARPES), or both.

\begin{figure}[b]
\includegraphics[width=8cm]{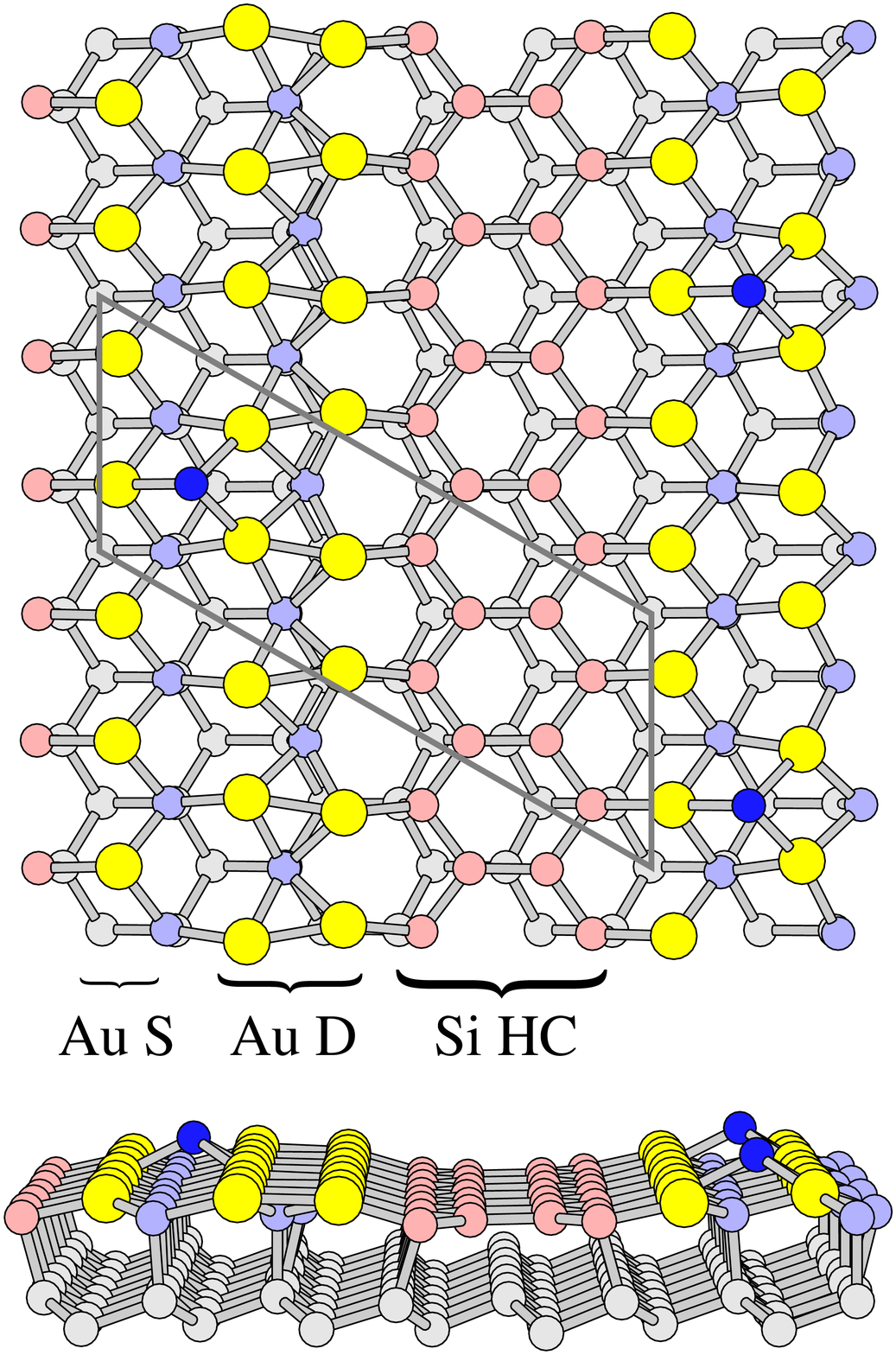}
\caption{(color online). Proposed structure of Si(111)-(5$\times$2)-Au
  with gold coverage equal to 0.6 monolayer. Large yellow circles are gold,
  small circles are silicon.  The surface layer consists of a gold
  single row (S), gold double row (D), and silicon honeycomb chain
  (HC).  The surface energy is minimized when this surface is
  decorated by silicon adatoms (dark blue) with 5$\times$4
  periodicity, as shown. In the presence of adatoms the 5$\times$1
  periodicity of the underlying substrate spontaneously doubles to
  5$\times$2 (gray outline) due to dimerization within the gold double
  row.  \label{model}}
\end{figure}

In this paper we propose a new structural model for
Si(111)-(5$\times$2)-Au that is fully consistent with all experimental
data to which we have compared. The model is similar to one proposed
by Erwin in 2003,\cite{erwin_phys_rev_lett_2003a} but is modified to
be consistent with a recently revised value of the gold
coverage.\cite{barke:155301} The modifications, although seemingly
minor, for the first time bring the predictions of the model---for STM
and ARPES as well as other phenomena---into excellent agreement with
experiment. More importantly, the new model opens the door to a more
fundamental physical understanding of Si(111)-(5$\times$2)-Au based on
its detailed atomic structure.

\section{Structural model}\label{structuralModel}

The model proposed here is shown in Fig.\ \ref{model}. The basic
structure is similar to one proposed several years ago in Ref.\
\onlinecite{erwin_phys_rev_lett_2003a} but there are several important
differences. For this reason it is useful to discuss both the
similarities and differences between the older model and
the one proposed here (hereafter the ``2003 model'' and the ``2009
model'').

The starting point for the 2003 model was the experimental observation
that 0.4 monolayer (ML) of gold induces a stable reconstruction of the
Si(111) surface. Using this coverage value the 2003 model was constructed
with four gold atoms per 5$\times$2 cell. As shown in Fig.\ 1 of Ref.\
\onlinecite{erwin_phys_rev_lett_2003a}, the gold atoms substitute for silicon
atoms in the topmost surface layer, forming two Au-Si chains oriented
along the [1$\bar{1}$0] direction. Adjacent to these two chains is a
silicon ``honeycomb chain,'' a thin graphitic strip of silicon that owes its
stability to a Si=Si double bond. The basic reconstruction just
described has 5$\times$1 periodicity. The experimentally observed
5$\times$2 substrate periodicity was argued to arise from a
row of silicon rebonding atoms that bridges a channel
between one of the Au-Si rows and the silicon honeycomb chain. A full row
of rebonding atoms overcoordinates some atoms, while a half-occupied
row, with 5$\times$2 periodicity, leaves some dangling bonds
unsaturated. The latter arrangement was argued to be energetically preferred
when the system is doped with extra electrons. These were supplied by
silicon adatoms adsorbed on top of the Au-Si chains.
 
The recently revised experimental determination of the gold coverage
as 0.6 ML obviously calls for a revised structural model as
well.\cite{barke:155301} The 2009 model accommodates the additional two
gold atoms per 5$\times$2 cell by replacing the half-occupied
rebonding row of silicon atoms from the 2003 model with a full row of
gold atoms. The other two building blocks of the 2003 model---the
silicon honeycomb chain and the adsorbed silicon adatoms---are
unchanged in the new model.  The total coverage of silicon atoms in
the top layer of the 2009 model varies, according to the coverage of
adatoms, between 1.20 ML (for the undecorated surface) and 1.25 ML
(for saturation coverage). These values are consistent with the range
of experimentally determined silicon coverage, 1.1 to 1.3 ML (Ref.\
\onlinecite{tanishiro1990a}), 1.23$\pm$0.003 ML (Ref.\
\onlinecite{seifert_dissertation_2006a}), and 1.3$\pm$0.1 ML (Ref.\
\onlinecite{chin2007a}). 
The 2009 model is energetically more favorable than the 2003
model: the theoretical surface energy is 17.1 meV/\AA$^2$
lower in the Au-rich limit, in which the chemical potential 
is taken as the bulk energy per atom.\cite{chempotAu}

Much of the rest of this paper will focus on the interesting role played by
the new row of gold atoms. But it is worth pausing briefly to
place Si(111)-(5$\times$2)-Au in the larger context of other
metal-induced reconstructions of Si(111). It is now known that a great
variety of metal adsorbates induce closely related reconstructions
based on the silicon honeycomb chain. But the details differ, sometimes
with unexpected consequences. Easily ionized adsorbates such as
alkalis, alkaline earths, and some rare earths form a family of
``honeycomb-chain channel'' (HCC) reconstructions in which the
adsorbates occupy channels between adjacent silicon honeycomb
chains.\cite{collazo-davila_phys_rev_lett_1998a,
lottermoser_phys_rev_lett_1998a,
erwin_phys_rev_lett_1998a} Each
adsorbate is three-fold coordinated by silicon atoms in the honeycomb
chain, which are too far away (3.0 \AA) to form covalent bonds,
consistent with a picture of ionic charge donation and electrostatic attraction.
The
interactions between adsorbates are also electrostatic, but
repulsive.\cite{erwin_surf_sci_2005a} Within this family of HCC reconstructions the adsorbate
coverage is determined by a simple electron-counting rule first
proposed by Lee {\it et al.}\cite{lee2001a,lee2003a}~and later
generalized 
by Battaglia {\it et al}.\cite{battaglia2007a,battaglia:075409}

Adsorbates that are less ionic, such as silver and gold, also form
reconstructions based on the silicon honeycomb chain but the role of
the adsorbate is more interesting. Silver adsorbates occupy the channel of
a HCC-like reconstruction,\cite{collazo-davila_phys_rev_lett_1998a}
but the stronger interaction between silver and silicon leads to a
preference for two-fold coordination of silver by silicon, at a much
reduced distance of 2.6 \AA.  The two-fold coordination brings silver
adsorbates sufficiently close to each other to allow pairing into silver
dimers. Chuang {\it et al.}\cite{chuang2008a} and Urbieta {\it et
al.}\cite{urbieta2009a} showed that the phase of this pairing
alternates between adjacent channels, modulating the periodicity of
the basic 3$\times$1 HCC reconstruction to a $c(12\times 2)$ variant.

Returning now to Si(111)-(5$\times$2)-Au, it appears that the role of
the adsorbate is still more complex. Gold is very reactive on silicon
surfaces.\cite{doremus2001a} This reactivity is already evident in the Au-Si rows of
Fig.\ \ref{model}. Within these rows gold completely substitutes for the
top layer of the surface silicon bilayer, with each gold atom covalently
bonded to three silicon atoms. Likewise, each silicon atom at the edge of the silicon honeycomb chain
forms a covalent bond to a gold atom. The Au-Si bond
lengths, both within the Au-Si rows and bonded to the silicon honeycomb chain,
are 2.5 \AA, smaller than for any other adsorbate studied. 

The most 
interesting aspect of the 2009 model is the behavior of the
gold double row labeled ``Au D'' in Fig.\ \ref{model}. As shown in
the figure, the equilibrium geometry of this double row is
dimerized. Period doubling was also part of the 2003 model, but its
origin---the half-occupied rebonding row---was simpler.  In the 2009
model the dimerization occurs only in the presence of silicon adatoms, or
when the surface is doped with extra electrons. In Sec.\ \ref{energetics} we
show that these two scenarios are largely equivalent. We further
demonstrate that in the presence of adatoms the dimerization is
driven by an unusual ``double'' Peierls mechanism, in which the distortion opens {\it
two} gaps in the band structure of the undistorted 5$\times$1
substrate.

\section{Methods}
\subsection{Theoretical methods}

First-principles total-energy calculations were used to determine
equilibrium geometries and relative energies of the basic
model and its variants.  The calculations were performed in a slab
geometry with four layers of Si plus the reconstructed top surface
layer and a vacuum region of 8 \AA. All atomic positions were relaxed,
except the bottom Si
layer and its passivating hydrogen layer,
until the largest force component on every atom was below 0.01 eV/\AA. Total energies and
forces were calculated within the PBE generalized-gradient
approximation to density-functional theory (DFT) using
projector-augmented-wave potentials, as implemented in {\sc vasp}.
\cite{kresse_phys_rev_b_1993a,kresse_phys_rev_b_1996a}  The
plane-wave cutoff for all calculations was 250 eV.

The sampling of the surface Brillouin zone was chosen according to the
size of the surface unit cell and the relevant precision requirements.
For example, the dependence of the total energy on dimerization (Fig.\
\ref{energy-vs-dimerization}) was calculated using a 5$\times$4 unit
cell and 2$\times$2 zone sampling, with convergence checks using
4$\times$4 sampling. The dependence of the relative surface energy on
silicon adatom coverage (Fig.\ \ref{energy-vs-coverage}) requires
greater precision because the energy variations are smaller.  Hence
these surface energies were calculated using a 5$\times$8 unit cell
(to allow an adatom coverage of 1/8) and 8$\times$4 sampling, with
convergence checks using 12$\times$6 sampling. Finally, the
potential-energy surface for adatom diffusion (Fig.\ \ref{diffusion})
was calculated using a 5$\times$4 unit cell and 2$\times$2 zone
sampling.

Simulated STM images (Fig.\ \ref{stm}) were calculated using the method of
Tersoff and Hamann.\cite{tersoff_phys_rev_b_1985a}  For the
filled-state image we integrated the local density of states (LDOS)
over a chosen energy window of occupied states up to the Fermi level;
for the empty-state image the integration was over unoccupied states
starting at the Fermi level.  The simulated STM topography under
constant-current conditions was obtained by plotting the height at
which the integrated LDOS is constant.

\subsection{Experimental methods}

Silicon wafers (from Virginia Semiconductors) were degassed for
several hours at 700 $^\circ$C before flashing at 1250 $^\circ$C for a
few seconds. A rapid cool-down to 850 $^\circ$C was followed by slow
cooling to room temperature. An important prerequisite for obtaining
well-defined row structures was a flexible mount of the samples that
prevented strain from building up during high-temperature flashing.
Gold was evaporated from a Mo wire basket.
The pressure was kept below
5$\times$10$^{−10}$ mbar throughout the sample preparation.
All STM measurements were carried out at room temperature with low
tunneling currents ($\le$50 pA).

Band dispersions were obtained using a Scienta 200
spectrometer with $E,\theta$ multidetection and an energy resolution of 20
meV for electrons and 7 meV for photons. We used $p$-polarized synchrotron
radiation at a photon energy $h\nu$ = 34 eV, where the cross section of
silicon surface states has a maximum relative to the bulk states.
The data are for a sample
temperature below 100 K.

\section{Energetics}\label{energetics}

This section addresses three issues related to the energetics of
Si(111)-(5$\times$2)-Au. We consider first the energetics of
dimerization, and show that dimerization occurs naturally within the
2009 model when silicon adatoms or extra electrons are present.  Next
we examine how the surface energy varies as a function of silicon
adatom coverage; we find that the energy is minimized for the coverage
1/4 shown in Fig.\ \ref{model}.  Finally, we show that the model
explains the existence of a localized defect structure seen even on
carefully prepared surfaces.

\subsection{Dimerization of the substrate}

The model shown in Fig.\ \ref{model} has 5$\times$4 periodicity,
because silicon adatoms decorate the surface in a 5$\times$4
arrangement.  But it is clear from the figure that the underlying
substrate (that is, ignoring the adatoms) can be understood more
simply as a 5$\times$1 reconstruction whose periodicity is doubled,
along the chain direction, to 5$\times$2 by dimerization within the Au
double row. Understanding the nature and origin of this dimerization
is important for explaining many experimental aspects of
Si(111)-(5$\times$2)-Au, including the fine details of STM imagery,
data from ARPES, the observation of nanoscale phase
separation,\cite{kirakosian_surf_sci_2003a,kirakosian_phys_rev_b_2003a,
  mcchesney_phys_rev_b_2004a,yoon_phys_rev_b_2005a,choi_phys_rev_lett_2008a}
the diffusion of silicon adatoms on the
surface,\cite{hasegawa_phys_rev_b_1996a,bussmann_phys_rev_lett_2008a}
and the existence and motion of domain walls within the Au-Si
rows.\cite{kang_phys_rev_lett_2008a}

\begin{figure}
\includegraphics[width=8cm]{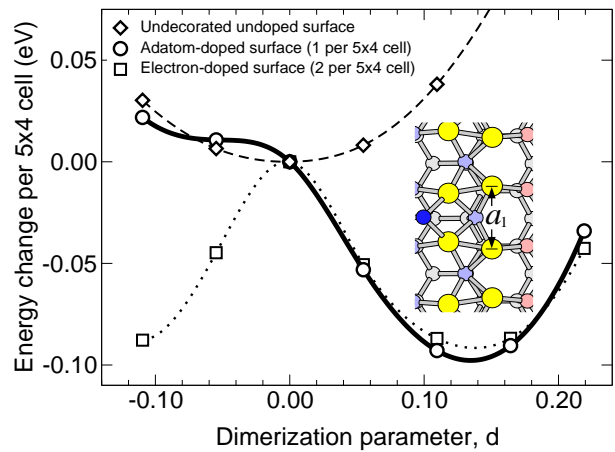}
\caption{(color online). Variation of the total energy 
as a function of dimerization along the chain direction. 
The dimensionless dimerization parameter is
  $d=(a_1-a_0)/a_0$, where $a_1$ is shown in the inset
  and $a_0$ is the surface lattice constant.
Total energies were calculated with full relaxation for each
constrained value of $d$.
\label{energy-vs-dimerization}}
\end{figure}

We begin by defining more precisely the nature of the dimerization.
The inset to Fig.\ \ref{energy-vs-dimerization} shows a detail of the
gold double row near a silicon adatom. Within this double row the gold
atoms have a ladder-like arrangement, with Au-Au bonds as the rungs.
In the absence of adatoms (or extra electrons) these rungs are all
parallel to each other, with a spacing equal to the silicon surface
lattice constant $a_0$. When adatoms decorate the surface in the
5$\times$4 arrangement shown in Fig.\ \ref{model} the rungs rotate
away from their parallel alignment. This rotation occurs almost
completely within the (111) surface plane, and the rungs are quite
rigid: the Au-Au bond length (2.94 \AA) changes by less than 1\%. This
rigidity is not surprising, because the Au-Au bond length is already very
close to the bulk gold bond length (2.88 \AA). The sign of the
rotation alternates along the chain direction. Hence the dimerization
can be viewed as an antiferrodistortive instability.

To quantify the dimerization one could, of course, use the angle of
rotation of the Au-Au rungs. We choose instead a more physically
transparent measure: the distance $a_1$ between gold atoms on the side
of the double row adjacent to the silicon honeycomb chain, as labeled in
Fig.\ \ref{energy-vs-dimerization}. A dimensionless dimerization
parameter can then be defined, $d=(a_1-a_0)/a_0$. For the 5$\times$4
arrangement of adatoms shown in Fig.\ \ref{model}, the equilibrium
dimerization parameter is $d_{\rm eq}=0.14$.

Some insight into the origin of the dimerization may be obtained by
computing the DFT total energy $E_t$ as a function of $d$
while relaxing all other degrees of freedom. The results are shown in
Fig.\ \ref{energy-vs-dimerization} for three variants of the basic
model: the
undecorated and undoped surface (without adatoms or extra electrons); 
the adatom-doped 5$\times$4 surface of Fig.\ \ref{model}; and
the undecorated surface doped with two extra electrons per 5$\times$4
cell. For the undecorated
undoped surface the minimum is at $d=0$, indicating that dimerization
is not stable. For the adatom-doped surface there is a single minimum in the
energy, as expected, at $d=+0.14$. For the electron-doped surface the behavior of $E_t(d)$
is nearly indistinguishable, for positive $d$, from that of the
adatom-doped surface. {\it This similarity strongly suggests that each
  silicon adatom dopes two electrons to surface states.} More substantive
evidence for this conjecture is found in the electronic structure of
these variants, as
we show below in Sec.\ \ref{Electronicstructure}.  

The behavior of $E_t(d)$ for negative $d$, however, is very different.
This is because adatom-doping strongly breaks the 5$\times$2
symmetry of the surface, while electron-doping does not. This suggests
another role for the silicon adatoms: {\it to pin the phase of the dimerization
such that $d>0$ at the position of the adatom.} We will return to this
role in Sec.\ \ref{diffusionofsiliconadatoms} when considering the diffusion of adatoms.

\begin{figure}
\includegraphics[width=8cm]{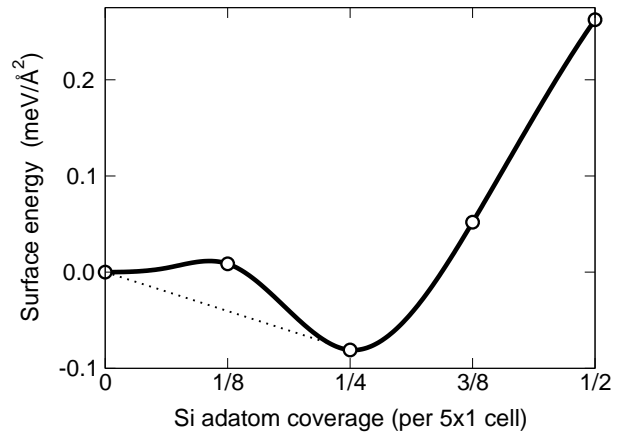}
\caption{Theoretical surface energy as a function of silicon adatom
  coverage. Energies were calculated at the labeled coverages; the
  interpolating curve is a guide to the eye.
  Adatoms occupy their preferred binding sites as
  shown in Fig.\ \ref{model}. The surface energy is lowest for
  coverage 1/4, corresponding to the
  fully saturated 5$\times$4 arrangement of adatoms.
  The dotted line highlights a local maximum 
  of the energy, and shows that a surface 
  with coverage less than 1/4 will phase separate
  into a mixture of empty and 1/4-covered regions, as found experimentally.
  \label{energy-vs-coverage}}
\end{figure}

\subsection{Adatom coverage and phase separation}

The preceding discussion and calculations were based on the assumption
that silicon adatoms decorate the surface in a 5$\times$4 arrangement,
equivalent to a coverage of 1/4 adatom per 5$\times$1 cell. 
This phase can be achieved experimentally by depositing silicon
onto the surface at temperatures around 300 $^\circ$C until saturation
is reached.\cite{bennewitz_nanotechnology_2002a} But this
saturated phase is metastable: annealing causes half of the silicon adatoms to diffuse
away. The resulting equilibrium phase is not uniform, but instead
exhibits patches with local adatom coverage of
1/4 interspersed with patches of undecorated surface;
the global average adatom coverage is close to
1/8.\cite{kirakosian_surf_sci_2003a,kirakosian_phys_rev_b_2003a,
mcchesney_phys_rev_b_2004a,yoon_phys_rev_b_2005a,choi_phys_rev_lett_2008a}
This experimental result poses two questions for theory. (1) What coverage of
silicon adatoms minimizes the calculated surface energy? (2) Does the
observed patchiness of the adatom distribution arise from an instability
toward phase separation?

To compare the energies of phases with different silicon adatom
coverage, we compute the relative surface energy
\begin{equation}
E_s = E_t(N_{\rm Si}) - N_{\rm Si}\;\mu_{\rm Si},
\label{relativesurfacenergy}
\end{equation}
where $E_t(N_{\rm Si})$ is the total energy of a surface unit cell
containing $N_{\rm Si}$ silicon atoms (including adatoms), and
$\mu_{\rm Si}$ is the silicon chemical potential. Thermodynamic equilibrium between
the surface and the bulk requires $\mu_{\rm Si}$ to be the
energy per atom in bulk silicon. Although we do not explicitly consider 
the free energy at finite temperature, configurational entropy effects
may indeed play a role; see below.

Within this formalism, adatoms are thermodynamically stable only if
their presence lowers the surface energy of the undecorated surface.
Equation \ref{relativesurfacenergy} shows that this requirement
implies that the adsorption energy per adatom must be greater than the
silicon cohesive energy (5.4 eV within DFT/PBE). Adsorption energies
below this threshold imply that adatoms may be temporarily metastable,
but not thermodynamically stable.  This is a very stringent
requirement, and one that is rarely satisfied in our experience (it
was not in the 2003 model). 

\begin{figure*}
\includegraphics[width=16cm]{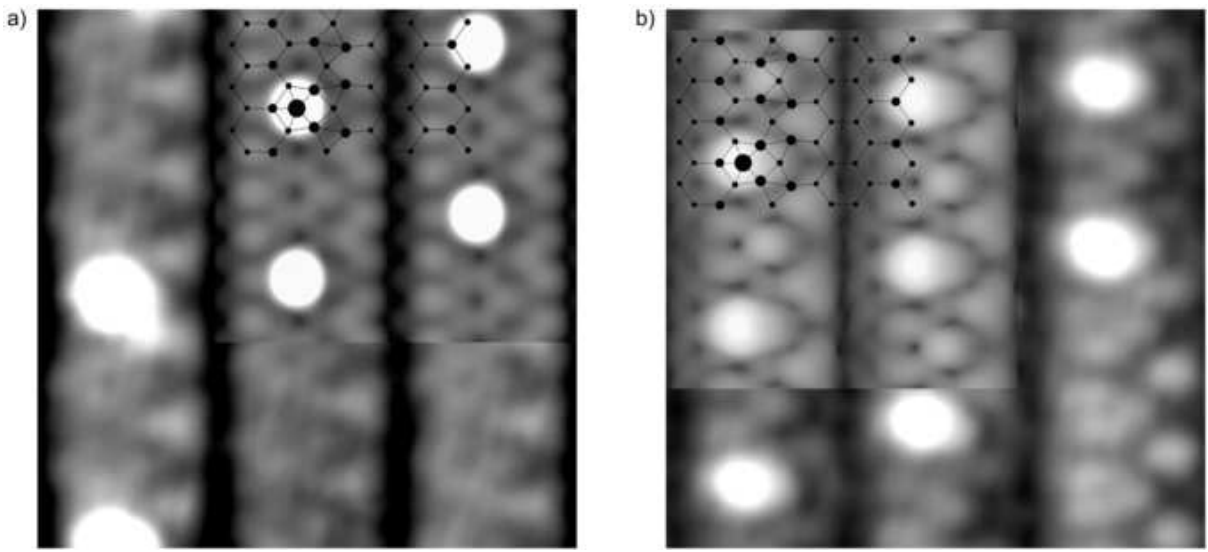}
\caption{Comparison of experimental and simulated (inset) STM images
  for Si(111)-(5$\times$4)-Au. (a)
  Filled states, $V=-$0.7 eV. (b) Empty states, $V=+$1.0 eV. The
  simulated images were obtained using slightly different voltages: $V=-$1.0 and
  $+$0.5 eV, respectively. Projected 
  positions are shown for silicon adatoms and top-layer silicon and gold atoms; the size of the circles indicates the
  height of the atoms.
  \label{stm}}
\end{figure*}

We calculated the relative surface energies for
five different silicon adatom coverages. To minimize numerical
uncertainties the same 5$\times$8
supercell was used for each coverage, and full relaxation 
performed in every case. The results are shown in
Fig.\ \ref{energy-vs-coverage}. The surface energy is minimized for a
coverage of 1/4. This is the coverage depicted in Fig.\ \ref{model},
and agrees well  with the experimentally observed local
coverage within adatom-covered patches. In Sec.\
\ref{Electronicstructure} we show why the value 1/4 is
special: at this adatom coverage---but at no other---the surface band
structure is fully gapped and hence favored because
occupied states can move down in energy.

We turn now to the second question posed above. Of the five coverages
considered here, the second most-favorable is not the adjacent 1/8 or
3/8 phase, but rather the undecorated ``empty'' surface. In other
words, the surface energy of the intermediate 1/8 phase is higher than
the average of the two endpoint phases, empty and 1/4. This result
suggests that a range of such intermediate coverages between 0 and 1/4
may have energies above the tie line, shown in Fig.\
\ref{energy-vs-coverage}, connecting the two endpoint phases.  {\it
  This implies that a surface prepared with adatom coverage 1/8 (and
  perhaps any intermediate coverage between 0 and 1/4) will
  phase separate into a mixture of empty and 1/4-covered
  regions.}  

Of course, this conclusion does not answer the related question of why
the observed global average coverage is close to 1/8. We hypothesize
that the 1/8 phase may be stabilized at finite temperature by the
entropy gained from occupying only half the adatom sites of the 1/4
phase.  We also do not address here the characteristic size of the
phase-separated regions; this would require analyzing the energy cost
of forming the phase boundary between the 0 and 1/4 phases.  Finally,
we leave for future analysis the possibility that the phase-separated
state benefits electrostatically from the charge transfer proposed by
Yoon {\it et al.}~to take place between the 0 and 1/4
phases.\cite{yoon_phys_rev_b_2005a}

\subsection{Low-energy defect structure}

In addition to the phase separation just discussed, STM images of
Si(111)-(5$\times$2)-Au sometimes reveal the presence of an occasional
``mirror-domain'' defect.\cite{yoon_phys_rev_b_2005a,seifert_dissertation_2006a} The
signature of this defect is a slight shift, by roughly 3 \AA\ to the
right in Fig.\ \ref{model}, of the bright protrusions that dominate
the STM topography. The intensity of the protrusions and the
position and shape of the underlying features are not affected.

A simple variation of the basic structural model accounts for the observed defects.
Starting from the structure shown in Fig.\ \ref{model}, one can
construct a mirror-image variant by reflecting the entire top
layer of atoms through the vertical plane bisecting the Si=Si double
bonds of the honeycomb chain. (This mirror plane was discussed in
Ref.\ \onlinecite{erwin_phys_rev_lett_1998a} and is shown in Fig.\ 1
therein.) The reflection leaves the honeycomb chain
unaffected but reverses the ordering and orientation of the Au-Si
rows.  The geometry of the rigidly reflected structure is already
extremely close (within a few hundreths of an \AA ngstrom) to the
fully relaxed geometry. The reflection shifts the silicon adatoms by
2.2 \AA\ to the right, consistent with the observed shift.  

The surface energy of the relaxed defect reconstruction is only
slightly higher, by 1.2 meV/\AA$^2$, than that of the original
reconstruction. From inspection of the atomic positions of the original and
defect models, we judge the energy cost of forming an interface
between the two phases to be quite small.  Thus it is plausible that
short sequences of this defect structure---manifested as a small
rightward shift of the silicon adatoms---should appear even on
carefully prepared surfaces, with minimal disruption to either the
primary reconstruction or the local adatom coverage.

\section{Comparison with STM}

Two decades of studies probing the Si(111)-(5$\times$2)-Au surface
with STM have created a richly detailed and consistent picture of its
topography and bias dependence.
\cite{baski_phys_rev_b_1990a,omahony_surf_sci_1992a,omahony_phys_rev_b_1994a,shibata_phys_rev_b_1998b,hasegawa_surf_sci_1996a,hasegawa_phys_rev_b_1996a,hasegawa_surf_sci_1996b,hasegawa_journal_of_vacuum_science__technology_a_1990a}
These data make possible a very stringent test for any proposed
structural model by comparing theoretically simulated STM images to
atomic-resolution experimental images. In this section we show that
simulated images based on the 2009 model are in excellent and detailed
agreement with recent experimental images. Moreover, the new model
resolves a small but irritating discrepancy found in comparisons
based on the earlier 2003 model.

The most pronounced features in STM imagery of the
Si(111)-(5$\times$2)-Au surface are the bright protrusions 
now well established as originating from the silicon adatoms. In
topographic maps these are generally the highest features in both filled- and
empty-state images. As discussed elsewhere 
\cite{mcchesney_phys_rev_b_2004a} and in the previous section, the
adatoms occupy a 5$\times$4 lattice.  Much detailed information is
contained in the imagery in between these lattice sites; the
topography of this substrate has
5$\times$2 periodicity. Of special interest is the registry of the
5$\times$4 adatom lattice and the 5$\times$2 substrate; the relative
alignment of these two lattices provides an important test for 
structural models. (We do not address here the lack of correlation
between the adatoms in different rows, previously discussed in
Ref.\ \onlinecite{kirakosian_phys_rev_b_2003a}.)

Figure \ref{stm} shows experimental and theoretical simulated STM
images for filled and empty states of the Si(111)-(5$\times$2)-Au
surface, in a region where the silicon adatom coverage is
1/4. The agreement between experiment and theory is excellent, and 
allows all of the main experimental features to be easily
identified. (1) The dark vertical channels separating the three main rows
shown in Fig.\ \ref{stm} arise from the doubly-bonded rungs of the
silicon honeycomb chains. (2) The bright protrusions are from the silicon
adatoms. (3) The triangular features with $\times$2 periodicity,
constituting the right edge of the rows, arise from the combination of
two contributions: a pair of outer gold atoms brought close by
dimerization (the apex of the triangle) and two silicon atoms at the
left edge of the honeycomb chain (the base of the triangle).  (4) In
the filled states, the left edge of the rows is not well resolved, and
appears to have either $\times$1 or very weak $\times$2 periodicity;
these spots arise from the silicon atoms at the right edge of the
honeycomb chain. (5) In the empty states, the $\times$2 triangular
features on the right edge of the rows alternate with {\bf V}-shaped
features that open to the left. These are a combination of two
contributions: a pair of dimerized inner gold atoms (the apex of the
{\bf V}) and Si-Au bonds that are slightly dimerized by their
proximity to the gold double row (the arms of the {\bf V}).

In the experimental images the alignment of the 5$\times$4 adatom
lattice and the 5$\times$2 substrate topography is as follows. The
bright protrusions are located slightly to the left side within the
main rows---except in the defect regions discussed in the previous
section, where
the protrusions are shifted to an equivalent location on the right
side of the row. This off-center location is accurately reproduced in
the simulated images based on both the original model and the defect model
(not shown).
Along the rows of the experimental images, the 5$\times$4 protrusions are
symmetrically straddled by the 5$\times$2 triangular and {\bf
  V}-shaped features of the substrate. Fig.\ \ref{stm} shows that this
symmetric registry is properly reproduced in the simulated images---resolving
a problem with the 2003 model first pointed out by Seifert
whereby the alignment of
the two lattices was asymmetric.\cite{seifert_dissertation_2006a} 

\section{Electronic structure}\label{Electronicstructure}

Photoemission studies of Si(111)-(5$\times$2)-Au began in the
mid-1990s and continue today.\cite{collins_surf_sci_1995a,
  okuda_j_electron_spectroscopy_and_rel_phenom_1996a,
  okuda_appl_surf_sci_1997a,
  hill_phys_rev_b_1997a,hill_appl_surf_sci_1998a,
  losio_phys_rev_lett_2000a, altmann_phys_rev_b_2001a,
  himpsel_journal_of_electron_spectroscopy_and_related_phenomena_2002a,
  zhang_phys_rev_b_2002b,mcchesney_phys_rev_b_2004a,
  choi_phys_rev_lett_2008a} These studies have led to important
insights into the electronic structure of this complicated surface
and provide tests complementary to STM for evaluating
structural models. In this section we address three aspects of the
electronic structure of Si(111)-(5$\times$2)-Au: the detailed
mechanism that drives the period doubling discussed in Sec.\ \ref{energetics};
a comparison of the theoretical band structure to angle-resolved
photoemission data; and an explanation of which specific surface
orbitals give rise to the observed band structure.

\begin{figure*}
\includegraphics[width=17cm]{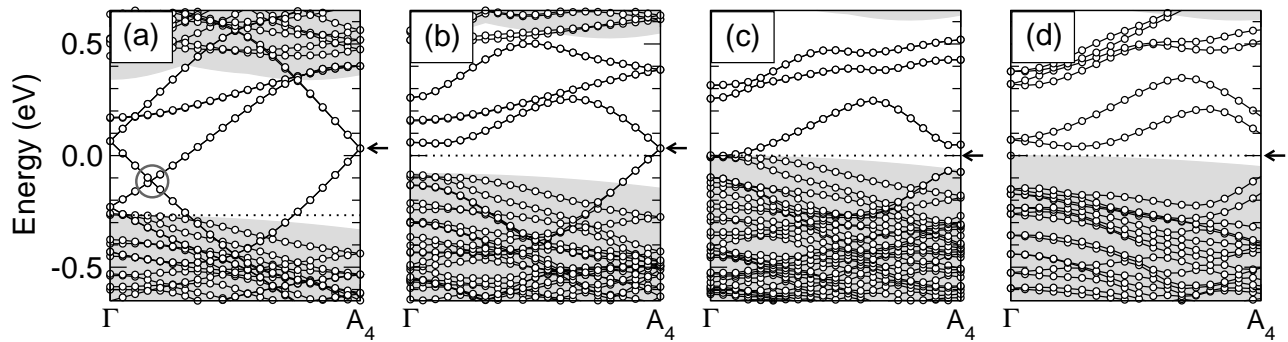}
\caption{Calculated band structure of Si(111)-Au in four different
  scenarios. (a) 5$\times$1 undecorated undoped surface. (b)
  5$\times$2 electron-doped surface (2 electrons per 5$\times$4 cell).
  (c) 5$\times$4 adatom-doped surface (1 adatom per 5$\times$4 cell);
  (d) 5$\times$4 adatom-doped surface with spin-orbit coupling
  included. In each panel the bands are plotted in the same 5$\times$4
  surface Brillouin zone.  The Fermi levels in panels (b), (c), and
  (d) are set to zero, and the zone-boundary degeneracies in (a) and
  (b) are aligned in order to highlight the evolution of the bands
  with doping.  Projected bulk silicon bands are shown in gray.  The
  formation of a hybridization gap at a band crossing (circled) and
  the opening of a gap at the zone boundary (arrows) are 
    indicated.  The observed surface is phase-separated into
    undecorated 5$\times$2 regions and adatom-doped 5$\times$4
    regions in equal proportion.
\label{bandStructures}}
\end{figure*}

\subsection{Origin of substrate period doubling}\label{Originofsubstrateperioddoubling}

In earlier sections the geometry and energetics of the period
doubling was explored without any consideration of the underlying
mechanism. Here we suggest an explanation, based on the 
electronic structure, for why the 5$\times$1 substrate dimerizes to
5$\times$2 in the presence of silicon adatoms or extra electrons.

We begin by considering a variant of the full 5$\times$4 model of Fig.\
\ref{model} in which the silicon adatoms are absent. When this surface is
relaxed within DFT the dimerization vanishes and hence the periodicity
reverts to 5$\times$1. The theoretical band structure for this surface is
shown in Fig.\ \ref{bandStructures}(a) for wave vectors parallel to
the chain direction and energies near the projected band gap. The
bands are plotted in the Brillouin zone of the full 5$\times$4 model
so that comparison with the bands of the full model can later be made. The folding
of the 5$\times$1 bands into the 5$\times$4 zone creates degeneracies
at the zone center $\Gamma$ and the 5$\times$4 zone boundary
A$_4$; the most important zone-boundary degeneracy is marked by an
arrow. The folding also creates band crossings, the most important of
which (circled) is inside the band gap, about one-fourth of the way
from $\Gamma$ to A$_4$. The Fermi level is very close to the top of
the valence band and the system is metallic.

Next we consider how this band structure changes when we dope the
surface with extra electrons. It was demonstrated in Fig.\
\ref{energy-vs-dimerization} that a doping level of 2 electrons per
5$\times$4 cell leads to stable 5$\times$2 dimerization, with
distortion energetics nearly identical to that from silicon adatoms at
1/4 coverage. Figure \ref{bandStructures}(b) shows the band structure
from this electron-doped surface. The electron doping has two
important consequences: the antiferrodistortive dimerization creates a
large hybridization gap at the band crossing of the the undoped
surface, and the Fermi level is pushed into this gap. The band
degeneracy at the 5$\times$4 zone boundary remains intact (see arrow),
because the electron-doped surface has perfect 5$\times$2 periodicity.

In the presence of silicon adatoms at 1/4 coverage this last
degeneracy is lifted. The last two panels of Fig.\
\ref{bandStructures} show the bands calculated at two levels of
theory: (c) scalar relativistic; (d) fully relativistic with
spin-orbit coupling. (We will show in the next section that the gold
character of these surface bands is substantial, hence the
spin-orbit splitting is large, about 0.2 eV.) Panel (d) shows that the
system now develops a full gap.  At 1/4 adatom coverage the
Fermi level falls just inside this gap, making the system insulating.

To summarize, we find that silicon adatoms at coverage 1/4 
create a multiband metal-insulator transition on Si(111)-(5$\times$2)-Au.
The first (electronically induced) gap arises from band hybridization
originating from dimerization and the resultant lowering of symmetry. The second
(adatom induced) gap arises from the 5$\times$4 potential of the adatoms,
which lifts the degeneracy at the 5$\times$4 zone boundary A$_4$.

\begin{figure}
\includegraphics[width=8cm]{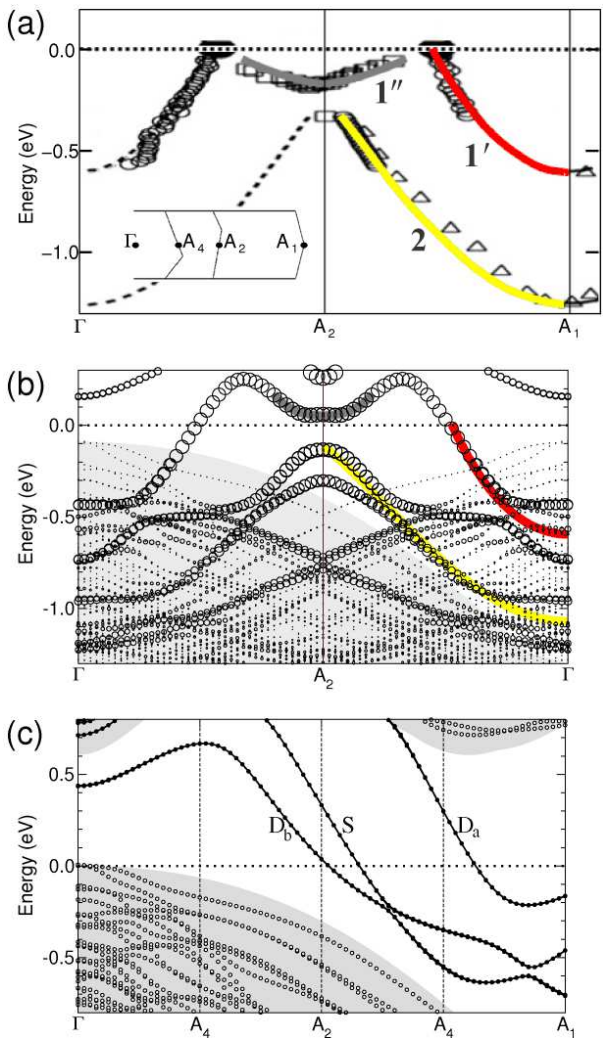}
\caption{(color online) Comparison of angle-resolved photoemission
  data to the theoretical band structure. (a) Photoemission-derived
  band dispersion reproduced from Ref.\
  \onlinecite{mcchesney_phys_rev_b_2004a}.  Band ${\bf 1'}$ (red) has
  strong 5$\times$2 periodicity, band ${\bf 2}$ (yellow) has mainly
  5$\times$1 periodicity, and band ${\bf 1''}$ (gray) has 5$\times$4
  periodicity and becomes more intense with increasing silicon adatom
  coverage.\cite{choi_phys_rev_lett_2008a} 
Inset: Brillouin zones for
5$\times$1, 5$\times$2, and 5$\times$4 surface unit cells.
(b) Theoretical band
  structure of the 5$\times$2 electron-doped surface. The diameter of
  each circle is proportional to the contribution from surface atoms
  in the Au-Si chains of Fig.\ \ref{model}.  (c) Theoretical band
  structure of the  hypothetical undecorated undoped 5$\times$1
    surface; these are to be compared with the folded bands in panels (a) and
    (b).  The labeled surface bands originate from gold and silicon
  orbitals in the single (S) and double (D) rows marked in Fig.~1. The
  double row leads to two bands, consisting of bonding (D$_{\rm b}$)
  and antibonding (D$_{\rm a}$) combinations of orbitals.
  \label{arpes}}
\end{figure}

\subsection{Comparison with angle-resolved photoemission}

The above discussion focused on bands above the Fermi level. Now we
turn to the occupied states, where we can make direct comparison to
experimental data.  

The experimental band structure is highly one-dimensional near the
Fermi level, and becomes gradually more two-dimensional for lower
energies; here we limit our discussion to the one-dimensional
dispersion along the chain direction.  
The results of our ARPES
studies of Si(111)-(5$\times$2)-Au are summarized in Fig.\
\ref{arpes}(a), which is reproduced from an earlier publication.
\cite{mcchesney_phys_rev_b_2004a}
By combining momentum- and
energy-distribution curves from several Brillouin zones, three bands
can be identified: ${\bf 1'}$, ${\bf 1''}$, and ${\bf 2}$. (These
labels are used by analogy to comparable bands at stepped Si(111)
surfaces.\cite{crain_phys_rev_b_2004a})  Figure \ref{arpes}(a) shows
the bands in the repeated-zone scheme of a surface with 5$\times$1
periodicity, for which A$_1$ is the zone boundary.  The 5$\times$2
period doubling introduces backfolded replicas of these bands, shown
as dashed lines.  The backfolded band ${\bf 1'}$ is indeed found where
expected, indicating that this band has strong 5$\times$2 character.
The backfolded replica of band ${\bf 2}$ is too weak to be
observed, indicated that it has mainly 5$\times$1 character.  Band
${\bf 1''}$ becomes more intense as the silicon adatom coverage is
increased, indicating that it has strong 5$\times$4
character.\cite{choi_phys_rev_lett_2008a}

It is difficult to compare directly the theoretical bands of the adatom-doped
surface to these experimental results, because the 5$\times$2
bands of Fig.\ \ref{arpes}(a) must be folded once more into the
Brillouin zone of the 5$\times$4 cell. This folding creates many bands
in a small energy interval and overly complicates the comparison
between theory and experiment.  We choose instead a simpler and
clearer approximate approach: to compare the experimental
5$\times$2 bands of Fig.\ \ref{arpes}(a) to the theoretical 5$\times$2
bands of the electron-doped surface with no silicon adatoms. In doing
so it must be kept in mind that the 5$\times$4 potential of the adatoms,
already seen to play an important role for the unoccupied bands, will
be absent.

Figure \ref{arpes}(b) shows the calculated bands for the 5$\times$2
electron-doped surface, plotted in the 5$\times$1 repeated-zone scheme
used in panel (a). (The projected bulk bands, however, are shown for
reasons of clarity in the extended-zone scheme.) The diameter of each
circle is proportional to the summed projections of the state onto gold and
silicon atoms in the single and double Au-Si chains of Fig.\
\ref{model}.  The solid colored curves represent our best effort to
match the three strongest surface bands to the ARPES bands. The
overall agreement for bands ${\bf 1'}$ and ${\bf 2}$ is excellent,
despite the complexity of the calculated bands even for this
simplified surface.  Note that in our interpretation, ${\bf 1'}$ and
${\bf 2}$ are not simple parabolic bands as depicted in Fig.\
\ref{arpes}(a). Instead, each comprises two or more bands and exhibits
several avoided crossings.  Further support for this interpretation
will be presented in the next subsection, where we discuss the orbital
origin of the bands.

The agreement appears less satisfactory for band ${\bf 1''}$. Although 
this band is correctly centered at the A$_2$ point, and the shallow
dispersion near that point reasonable, it is shifted rigidly up in
energy by 0.2 eV compared to experiment. We believe that this shift is
a spurious effect arising from the omission of silicon adatoms in the
calculation. Indeed, Choi {\it et al.}~have shown experimentally that
the ${\bf 1''}$ band shifts upward as the adatom coverage is reduced.
\cite{choi_phys_rev_lett_2008a}  For the range of coverages studied
(between 37 and 97\% of the saturation 1/4 coverage) the shift in
energy was linear. Extrapolating this result to a surface free of
adatoms, one naively expects an upward shift in energy of the ${\bf
  1''}$ band by 0.21 eV. Such a shift would bring the results of ARPES
and theory into excellent agreement for all three bands.

\subsection{Orbital origin of the bands}

We now make one last theoretical simplification by eliminating the
extra electrons. Upon relaxation the undoped surface is no longer
dimerized, and the periodicity reverts to 5$\times$1.  This
simplification is justified if the perturbation of the bands from the
dimerization and the extra electrons is small. In the limiting case of
zero dimerization the exact 5$\times$2 bands can be obtained from the
5$\times$1 bands by simple zone folding. Figure \ref{arpes}(c) shows
the calculated band structure for the simplified surface, plotted in
the conventional 5$\times$1 reduced-zone scheme. It is readily
apparent that the bands in (b) can indeed be accurately obtained from
those in (c) by first folding the 5$\times$1 bands about the
zone midpoint A$_2$, and then shifting the Fermi level
upward by the appropriate amount (0.3 eV) for electron doping of rigid
bands.

The advantage of the simplified 5$\times$1 band structure is that its
orbital origin is easy to analyze, because the bands fall almost
entirely in the projected gap and rarely cross. By examining
individual states in real space, we find that the middle band S
originates primarily from gold and silicon orbitals in the single
chain ``Au S'' in Fig.\ \ref{model}. The other two bands are more
complicated. They arise from the double chain labeled ``Au D.''  The
two bands are the low-lying bonding (D$_{\rm b}$) and higher-lying antibonding
(D$_{\rm a}$) combinations of orbitals of the gold atoms that constitute each
rung of the ladder (combined with orbitals on the connecting silicon
atoms).

By comparing the three panels of Fig.\ \ref{arpes} we can now understand
the orbital origin of the ARPES bands. Band ${\bf 1'}$ is the
simplest and corresponds to the antibonding D$_{\rm a}$ band. Band
${\bf 2}$ is a superposition of bands S and D$_{\rm b}$,
which are energetically for energies below the Fermi level. Band ${\bf
  1''}$ has more complicated origin. It arises from two
effects: 
strong rehybridization (related to the dimerization) around A$_2$ of
bands S and D$_{\rm b}$, and the energy shift discussed above
from the adatom potential.

We can also now better understand the two gaps created in Fig.\
\ref{bandStructures} by adatom doping. The orbital origin of the those
bands can now be assigned by noting, first, that the bands in Fig.\
\ref{bandStructures}(a) are identical to those in Fig.\
\ref{arpes}(c), folded twice along the labeled vertical lines. It
should be clear that the electronically induced gap illustrated in
Fig.\ \ref{bandStructures}(b) arises from hybridization of bands S and
D$_{\rm b}$, which cross near $\Gamma$ when folded into the 5$\times$4
zone. Likewise, the adatom-induced gap illustrated in Fig.\
\ref{bandStructures}(c) is created entirely within the D$_{\rm a}$
band at the second A$_4$ point, which folds into the A$_4$ zone boundary 
of the 5$\times$4 zone.

\begin{figure}
\includegraphics[width=8cm]{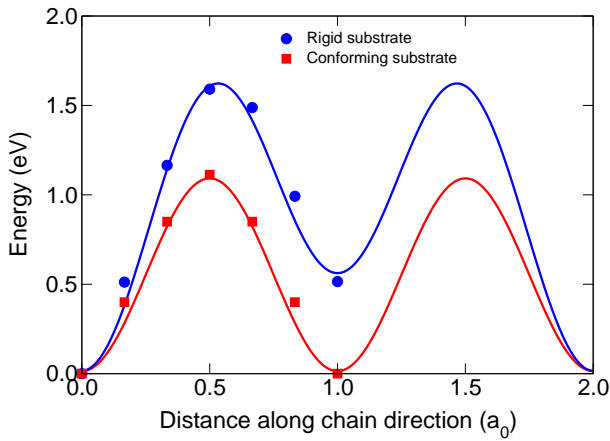}
\caption{(color online). Theoretical potential energy surfaces for the diffusion of a silicon
  adatom along the chain direction. Two 
approximations were considered for determining the diffusion
pathway. 
(1) Substrate atoms were fixed rigidly at
  their equilibrium positions attained when
  adatoms are in their lowest energy site (circles). (2)
  Substrate atoms were allowed to conform to their ``instantaneous'' equilibrium
  positions as the adatom diffused (squares). These two limiting cases give
  estimates for the activation barrier of 1.6 and 1.1 eV, respectively.
  The experimentally measured
  barrier is 1.24 $\pm$ 0.08 eV.\protect\cite{bussmann_phys_rev_lett_2008a}
  \label{diffusion}}
\end{figure}

\section{Diffusion of silicon adatoms}\label{diffusionofsiliconadatoms}

The silicon adatoms decorating Si(111)-(5$\times$2)-Au are not
immobile.  \cite{hasegawa_phys_rev_b_1996a} Sequences of STM images of
surfaces prepared with adatom coverage near the equilibrium value 1/8
show evidence for diffusion of adatoms along the Au-Si chain
direction, by a defect-mediated mechanism of unknown
origin.\cite{bussmann_phys_rev_lett_2008a} From measurements of the
mean-square displacement at different temperatures an effective
activation barrier of 1.24$\pm$0.08 eV was
extracted.\cite{bussmann_phys_rev_lett_2008a} In this section we
examine whether the structural model proposed in Fig.\ \ref{model} can
account for this diffusion barrier.

Although it is usually straightforward to calculate a barrier for the
diffusion of an atom on a well-defined surface,
Si(111)-(5$\times$2)-Au presents an interesting complication. Normally
one studies the diffusion of a single atom in a supercell representing
the clean surface, and relaxation of the surface around the
diffusing atom is allowed. But for Si(111)-(5$\times$2)-Au the
dimerization of the substrate is determined by the location (and
coverage) of the adatoms whose very motion is under study. Thus, if
the supercell is taken to be 5$\times$4 and the substrate is allowed
to relax, then the phase of the dimerization will track the position
of the adatom. This ``conforming substrate'' is physically
unrealistic and will underestimate the true diffusion barrier.

A more plausible scenario, in which the phase of the dimerization
remains fixed while the adatom diffuses, is difficult to implement
without constraints or a larger supercell. We take a simpler approach
and adopt a perfectly rigid 5$\times$4 substrate, which will
overestimate the true barrier. The true barrier must then be between
those of the conforming and the rigid substrate.

These two potential energy surfaces were calculated using a 5$\times$4
unit cell in which the projected
position of the adatom along the chain direction was constrained. For
the conforming scenario, all other degrees of freedom 
were relaxed. For the rigid scenario, the two remaining adatom
degrees of freedom were relaxed while all other atoms were 
were fixed at their original 5$\times$4 positions.

Figure \ref{diffusion} shows the two resulting potential-energy
surfaces.  For the conforming substrate the potential-energy surface
has 5$\times$2 periodicity and an activation barrier of 1.1 eV, while
for the rigid substrate the periodicity is 5$\times$4 and the
activation barrier is 1.6 eV. These two barriers nicely bracket the
experimental barrier of 1.24 eV.  More detailed
studies using realistic boundary conditions will likely play an
important role in unraveling the nature of adatom diffusion on this
surface.

\section{Outlook}

The structural model proposed here for Si(111)-(5$\times$2)-Au 
resolves one of the longest-standing unsolved reconstructions of the
silicon surface.  Predictions based on this model---for the dimerization of
the underlying substrate, for the saturation coverage of silicon adatoms, for
phase separation into adatom-covered and empty regions, for
detailed STM imagery, for electronic band structure, and
for the diffusion of adatoms---are in excellent
agreement with experiments. Moreover, the physical mechanisms
underlying many widely studied phenomena in this system have now been
elucidated. 

A number of other issues awaiting theoretical study can now be
directly addressed. These include, for example, the suggestion that
phase separation is accompanied by charge separation and the formation
of a Schottky barrier at the interface between adatom-covered and
undecorated regions; \cite{yoon_phys_rev_lett_2004a} the structure
and motion of the domain walls, within the Au-Si rows, that form when
the spacing between two neighboring adatoms is not commensurate with
the 5$\times$2 substrate; \cite{kang_phys_rev_lett_2008a} and the
exploration of the fundamental limits on using Si(111)-(5$\times$2)-Au
to store and manipulate digital information at densities comparable to that of
DNA.\cite{bennewitz_nanotechnology_2002a,kirakosian_surf_sci_2003a}

Finally, we anticipate a renewal of theoretical interest in
Si(111)-(5$\times$2)-Au as a physical realization of a nearly
one-dimensional metal. Most studies to date have used parametrized
models because a definitive structural model has not been
available.\cite{liu_nanotechnol_2008a} With a model in hand, the door
is now open to insights from new theoretical investigations as
well.

\section{Acknowledgements}

This work was supported by the Office of Naval Research, and by the
NSF under awards No. DMR-0705145 and DMR-0084402 (SRC).  S.C.E.
gratefully acknowledges many helpful discussions with Christoph
Seifert.  I. B. acknowledges support from the DAAD. Computations were
performed at the DoD Major Shared Resource Center at AFRL.


\begin{thebibliography}{10}
\bibitem{chempotAu}
A more realistic value for the gold chemical potential might decrease this
  energy difference, but is difficult to determine experimentally.
\end{thebibliography}

\end{document}